\documentclass[twocolumn,english,prl]{revtex4-1}
\usepackage{hyperref}
\usepackage[all]{hypcap} \usepackage[latin9]{inputenc}
\usepackage{amsmath}
\usepackage{amssymb}
\usepackage{graphicx}
\usepackage{babel}

\newcommand{\comment}[1]{}
\newcommand{\Amp}{\,\mathrm{A}}
\newcommand{\kAmp}{\,\mathrm{kA}}
\newcommand{\micron}{\,\mathrm{\mu m}}
\newcommand{\GeV}{\,\mathrm{GeV}}
\newcommand{\GVm}{\,\mathrm{GV/m}}
\newcommand{\metre}{\,\mathrm{m}}
\newcommand{\Joules}{\,\mathrm{J}}

\begin{document}

\title{Stable particle acceleration in co-axial plasma channels}

\author{Alexander Pukhov and John P. Farmer}

\affiliation{Institut f\"ur Theoretische Physik I, Universit\"at D\"ussseldorf, 40225
Germany}

\begin{abstract}
The attainable transformer ratio in plasma accelerators is limited by instabilities.  Using three-dimensional particle-in-cell simulations, we demonstrate that these can be controlled using a hollow plasma channel with a co-axial plasma filament.  The driver scatters electrons from the filament, and the slow pinch of the ions leads to a strong chirp of the effective betatron frequency, preventing beam breakup.  We demonstrate the monoenergetic acceleration of an electron bunch to $20\GeV$ over $4.4\metre$, achieving a transformer ratio of 10, an energy efficiency of 40\% and an emittance of $1.8\micron$.
%
\end{abstract}

\maketitle

Plasma wakefields \cite{esarey} offer a potential basis for
novel high-energy particle accelerators \cite{Joshi} due to the high
field gradients plasma can support. A driver excites a plasma
wake, which is in turn used to accelerate a trailing witness bunch.
The driver can be either an intense laser pulse \cite{bubble} or
a charged particle bunch \cite{Afterburner}.  Reaching high energies in both cases is challenging.  For a laser driver, the limitation is
due to dephasing, as the witness bunch travels faster than the laser
driver.  This can be overcome by using a staged acceleration\cite{multistage}.

In this work, we focus on the use of a particle driver.  One can use short - shorter than the plasma period
- drive bunches in quasi-linear \cite{Quasi-linear} or blow-out \cite{Blowout,Golovanov}
regimes. Alternatively, one may harness the self-modulation of longer
bunches in plasma \cite{Vph,AWAKE_Muggli_2017}. For the accelerating
medium, one may choose either uniform plasma \cite{Afterburner},
or a pre-formed plasma channel \cite{Andreev}. Each of these regimes
has its own particular advantages and drawbacks.

Perhaps the most promising accelerating scheme is that of the hollow
plasma channel \cite{Hollow}. A radially symmetric drive bunch in
a cylindrical channel does not generate any focusing or defocusing
fields, which would allow the use of a long drive beam, necessary
for high transformer ratios, and guarantee the conservation of the transverse
emittance of the witness \cite{Longqing}. Further, the accelerating field is
uniform across the hollow channel, allowing monoenergetic acceleration.
A high quality witness bunch is vital for a number of applications,
such as future high-energy colliders \cite{ILC} or XFEL machines
\cite{DESY}. Thus, hollow-plasma-channel acceleration appears the
perfect candidate for next-generation particle accelerators.

Unfortunately, hollow plasma channels suffer from a severe drawback
- the beam-breakup (BBU) instability \cite{BBU_Schroeder,Yakimenko_misaligned}.
As a charged bunch propagates in a hollow channel, plasma electrons
in the channel wall respond. The resulting space-charge results in
an attractive force between the bunch and the wall. The plasma response
increases as the bunch moves towards the wall, increasing the attractive
force. This instability manifests as a hosing of the beam: an oscillation
of the beam centroid along its length \cite{Whittum}. 
Ultimately, the bunch tail hits the wall of the channel and the bunch
is destroyed. The characteristic growth distance of the BBU instability
is sufficiently short that no significant energy exchange from the
driver to the wake can be achieved before the driver is lost. A similar
instability is observed in dielectric-based accelerators \cite{Single_Bunch_Limits_BBU}.

BBU is well known in other acceleration schemes. 
In conventional linear accelerators \cite{BBU_linacs} it is controlled through BNS-stabilization \cite{BNS}, in
which an energy chirp is applied to the bunch. The resulting head-to-tail
chirp in the betatron frequency breaks the resonance between the beam
and channel, suppressing the instability. The chirp must be consistent
with the focusing properties of the quadrupole guiding structure \cite{BNS_Stupakov},
and must be maintained over the whole accelerating/decelerating distance.
Recently, BNS stabilization has been successfully applied to dielectric-based
accelerating structures \cite{StrongBfocusing}. However, even with current state-of-the-art magnetic quadrupole
technology, offering field gradients on the order of $1\,\mathrm{T/mm}$, 
the attainable accelerating field is limited
to a few $100\,\mathrm{MV/m}$.  The presence of the instability places fundamental constraints
on the maximum accelerating field \cite{Zholents2}.

In the blowout regime of plasma wakefield acceleration, the BBU growth rate is significantly smaller than for a hollow channel due to the bubble geometry \cite{Huang_hose_bubble}.  This allows stabilization to be achieved by using
a drive beam with an initial energy spread \cite{Mehrling_hose_mitigation} or large transverse size \cite{Martinez_hose_intrinsic_stabilization}, or even through driver energy loss  \cite{Mehrling_hose_mitigation}.  These methods are either inapplicable or insufficient for the stabilization of the hollow channel.  However, the driver length in the blowout regime is necessarily limited by the bubble length, which places an upper limit on the transformer ratio, and so the efficiency.  In a hollow channel, the driver length is limited only by the BBU instability.

\begin{figure}
\includegraphics[width=8.5cm]{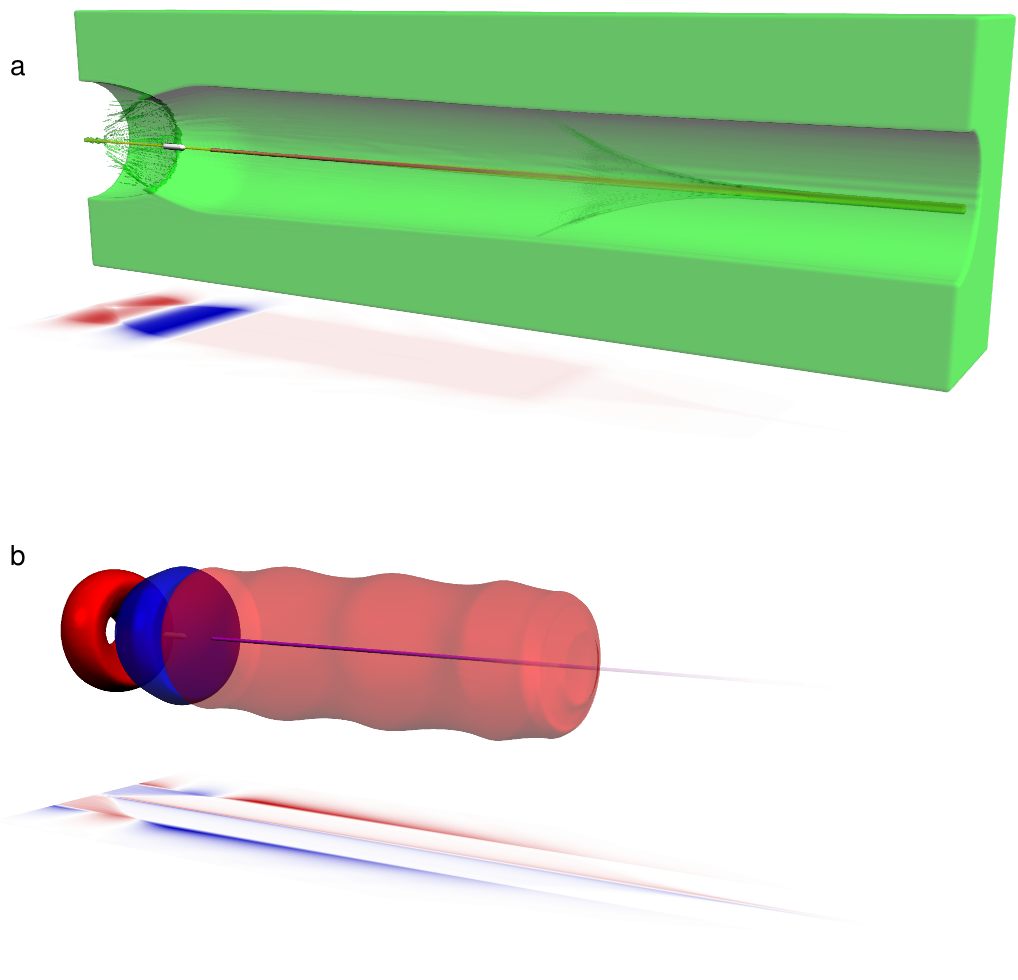} \caption{\label{fig:Channel_configuration}Configuration of the co-axial--channel accelerator.  a) A drive beam (purple)
propagates in a hollow channel, scattering the plasma electrons (green)
from the co-axial filament, leaving an ion column (yellow). The response of the bulk plasma generates a longitudinal electric field, shown as a 2D cut through the axis projected underneath.
This field allows the acceleration of a witness bunch (white) to energies
much higher than that of the driver. 
b) Field configuration, showing isosurfaces of the decelerating field
acting on the driver (translucent red, at $0.01E_{c}$, where $E_{c}=c\omega_{p}m/e=23\GVm$);
the accelerating field acting on the witness (blue, $-0.1E_{c}$);
and the field suitable for the acceleration of positrons (red, $0.1E_{c}$).  The driver and witness are also shown, for easy comparison to (a).
The projection underneath shows a 2D cut through the axis of the transverse
field $E_{y}-B_{z}$ acting on the driver and witness.}
\end{figure}

Here we show that stable acceleration in a hollow plasma channel can
be achieved through the inclusion of a thin co-axial plasma filament.
The accelerator configuration is shown in Fig.~\ref{fig:Channel_configuration}.
We assume the plasma density $n$ in the filament to be the same as
in the walls of the channel. The filament radius $r_{f}$ must be
small, so that $k_{p}r_{f}\ll1$. Here $k_{p}=\omega_{p}/c$ is the
characteristic plasma wave number, with $\omega_{p}=\sqrt{4\pi ne^{2}/m}$
is the corresponding electron plasma frequency. If we use an electron
drive bunch with a current $I_{d}\gg I_{A}\left(k_{p}r_{f}\right)^{2}$,
where $I_{A}=mc^{3}/e=17\kAmp$ is the natural current unit,
the transverse self-field of the driver will scatter the plasma electrons
from the filament. The remaining ion column will guide both the electron
driver and a negatively charged witness. Simultaneously, the ion column
will slowly pinch due to the high charge of the drive beam. The characteristic
pinch time is $\tau_{i}\sim\left(r_{f}/c\right)\sqrt{MI_{A}/ZmI_{d}}$,
where $M$ is the ion mass and $Z$ is the ion charge. As the ion
pinch begins at the head of the driver, the ion density, and so the
effective betatron frequency, increases along the length of the beam.
The large effective chirp guarantees the bunch stability through the BNS
mechanism \cite{BNS,BNS_Stupakov}, even for a monoenergetic driver.
This chirp is independent of the beam energy, allowing much larger chirp rates than can be achieved by tailoring the driver energy spread.  This makes the configuration ideal for exploiting the large acceleration gradients possible in a plasma accelerator.

To demonstrate stable acceleration in a co-axial channel, we use the
fully three-dimensional quasi-static particle-in-cell code qv3d, developed
on the platform of the VLPL code \cite{qv3d}. This makes possible simulations that would be infeasible using conventional simulation methods \footnote{Simulations were carried out in the light frame, $\xi=z/c-t$, with a resolution of $\Delta_{x}=\Delta_{y}=0.07/k_{p}$, $\Delta_{\xi}=0.1/k_p$, and a simulation box size of $15/k_{p}\times15/k_{p}\times50/k_{p}$.  The plasma response was modelled with a timestep $\Delta_{t}=1000/\omega_{p}$,
with the witness and drive beams subcycled with a timestep $\Delta_{t}=10/\omega_{p}$
in order to correctly resolve the betatron frequency. 4 particles
per cell were used for the bulk plasma, 64 for the drive beam and
plasma filament, and 4096 for the witness bunch.}.  We choose helium as
the background gas with an atomic density $n=5.7\times10^{16}\,\mathrm{cm^{-3}}$.
The hollow plasma channel has a radius $k_{p}r_{c}=3$, and the on-axis
plasma filament has a radius $k_{p}r_{f}=0.2$. In dimensional units,
these are $r_{c}=67\micron$ and $r_{f}=4.4\micron$.
The filament and channel walls are taken to be singly ionized. 
We do not discuss here how such a plasma configuration may best be 
achieved. The standard method to create a hollow channel is by
laser ionization \cite{Yakimenko_misaligned}. The co-axial filament could, for example, be ionized by a higher-order laser mode or even by the
self-field of the drive bunch \cite{Tarkeshian_self}.

Both the driver and witness have an initial particle
energy of $2\GeV$, a negligibly small energy spread, and an emittance of $1\micron$. The driver consists of two bunches.
The main driver has a ramped density profile, with a current increasing
from zero to $10\kAmp$ over $530\micron$,
and a Gaussian transverse profile with $\sigma_{\perp}=1.6\micron$.  This bunch duration is approximately equal to the characteristic pinch time for the ion column.

\begin{figure}
\includegraphics[width=8.5cm]{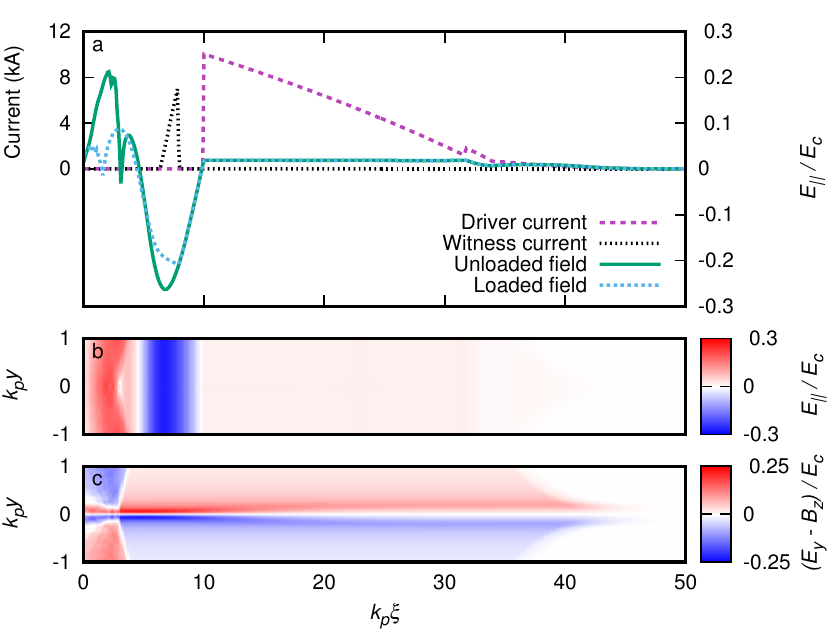} \caption{Beam profiles and wakefields.  a) Initial drive- and witness-beam currents and the on-axis longitudinal field,
plotted in the co-moving coordinate $\xi=z-ct$. The maximum decelerating
wakefield inside the drive beam is $E_{-}=0.019E_{c}$. The maximum
unloaded accelerating field reaches $E_{+}^{\mathrm{unloaded}}=-0.26E_{c}$,
corresponding to an unloaded transformer ratio of 13.6. When loaded
with the witness bunch, the accelerating field flattens to $<E_{+}^{\mathrm{loaded}}>=-0.2E_{c}\approx4.6\GVm$.  b) A 2D cut of the longitudinal field, $E_{||}$, through the axis.  c) A 2D cut of the transverse field $(E_y-B_z)/E_c$ through the axis.  
\label{fig:Accelerating-wakefield}}
\end{figure}

Such ramped density profiles minimize the decelerating field acting on the driver \cite{Bane}, allowing a larger transformer ratio.  However, the high-current driver used here modifies the equilibrium radius of the channel and induces a return current in the bulk plasma.  This results in a larger effective accelerator loss factor \cite{BBU_Schroeder} for higher beam currents, i.e.~a stronger coupling between the drive beam and the channel.  The optimal gradient for the main drive bunch is therefore sublinear.  We here make use of a logarithmic ramp profile $I(x)\sim\log(1+\alpha x/L)$, with $\alpha=0.057$, which corresponds to a first-order correction to the plasma response.

An additional nonlinear wake term arises due to the scattering of electrons from the on-axis filament.  This increases the decelerating field near the leading edge of the driver, reducing the transformer ratio obtained from commonly-used driver profiles, e.g.~the double-triangular bunch \cite{doubletriangular}.  We avoid this limitation through the use of a second drive bunch which precedes the main driver, scattering the filament electrons before the peak decelerating field is reached.  The leading bunch has a Gaussian rise, $\sigma_{||}=110\micron$, with a sharp cut to zero at its peak of $610\Amp$.  The transverse profile is the same as the main driver.  The two drive bunches partially overlap, with the start of the main driver $50\micron$ before the peak of the preceding bunch.

The combined current profile of the two drive bunches, and the resulting wakefield, is shown in Fig.~\ref{fig:Accelerating-wakefield}a.  
The maximum decelerating field is $E_{-}=0.019E_{c}$, where the critical
field $E_{c}=c\omega_{p}m/e=23\GVm$ for the chosen plasma
density. The field structure has an unloaded transformer ratio $T_{R}=13.6$. 
This value is $86$\% of the theoretical maximum for a main drive bunch of this length \hbox{\cite{Zholents1}} in a hollow channel of this radius \cite{BBU_Schroeder}.  We note that the decelerating field after the peak is flat to within $\pm2.3$\%.  Further optimization would require a higher-order treatment for the plasma response.

The leading edge of the witness is located $46\micron$
behind the rear edge of driver.   Its transverse profile is Gaussian with 
$\sigma_\perp=0.79\micron$, and a peak current of
$7\kAmp$ at its leading edge, decreasing linearly over
its $33\,\mu m$ length.  This density profile is chosen to correctly
load the wakefield.  The average accelerating field experienced by the witness $<E_{+}>=-0.2E_{c}$ corresponds to a loaded transformer ratio $T_{R}=10.2$.

The longitudinal ($E_{||}$) and transverse ($E_y-B_z$) fields near the channel axis are shown for the unloaded accelerator in Figs.~\ref{fig:Accelerating-wakefield}b and c.  The accelerating and focusing fields are The chirp in the transverse focusing field arising from the pinch of the ion column is immediately apparent.

\begin{figure}
\includegraphics[width=0.95\columnwidth]{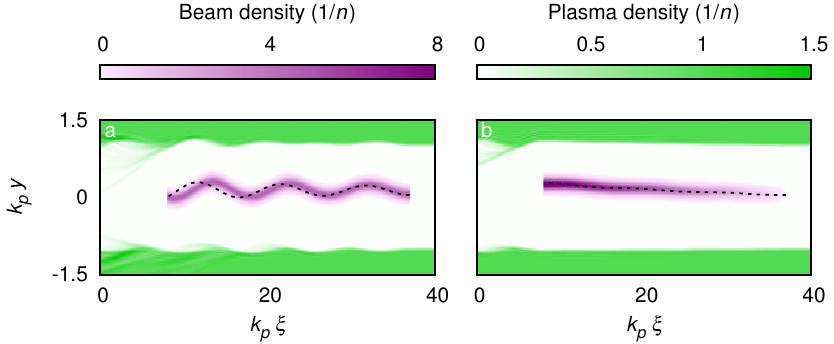} \caption{BBU instability.  Comparison of the qv3d code (color-map) with a semi-analytical model
(black dashed line) for BBU growth in the absence of a co-axial filament.
A flat-top (a) and ramped (b) drive beam of average current $200\Amp$
propagate $Lk_{p}=600$ in a channel of radius $r_{c}k_{p}=1$. The
instability is seeded by offsetting the driver by $0.05/k_{p}$ from
the channel axis. \label{fig:bench}}
\end{figure}

Without the co-axial plasma filament the driver rapidly becomes transversely
unstable. The BBU growth observed in the qv3d code is in good agreement
with analytical models, as seen in Fig.~\ref{fig:bench}, which compares
simulations with the numerical solution of Eq.~(13) from reference
\cite{BBU_Schroeder}. The results diverge as the plasma response becomes
nonlinear due to the limitations of the analytic model.
Without a plasma filament, simulations for the same parameters as used in Fig.~\ref{fig:Channel_configuration}
show the loss of the witness beam due to BBU over distances as short as $L_{\mathrm{BBU}}\approx4000k_{p}^{-1}\approx90\,\mathrm{cm}$, limiting the energy gain to $\sim400\,\mathrm{MeV}$.

\begin{figure}
\includegraphics[width=9cm]{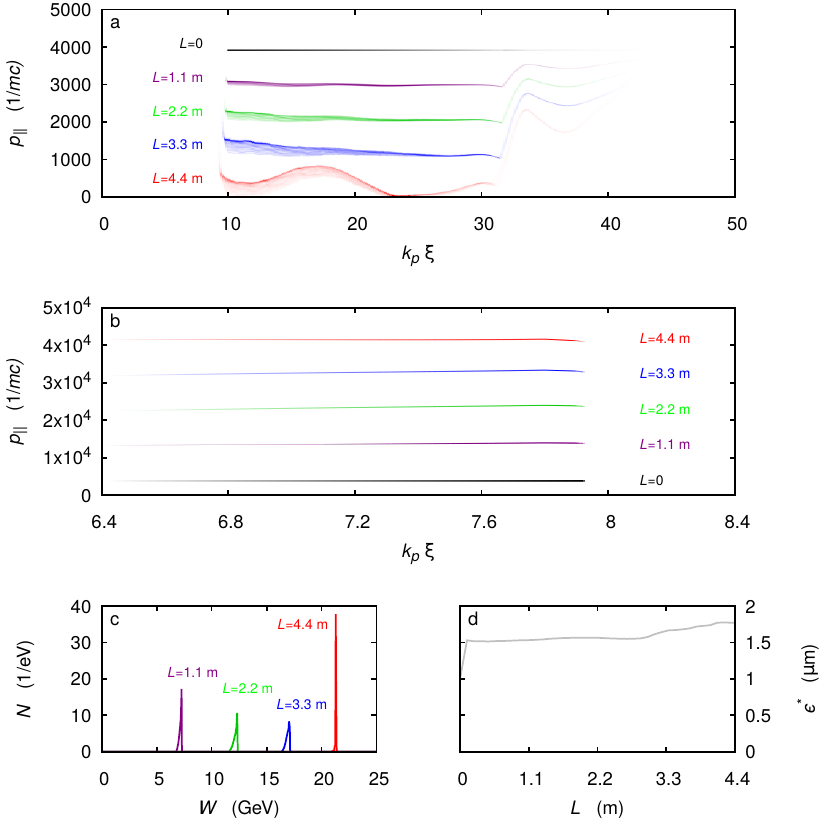}
\caption{Evolution of the drive and witness beams.  a)~Longitudinal phase-space density 
$(p_{||},\xi)$ of the driver after an acceleration length of $L=0$ (initial distribution), 
$1.1$, $2.2$, $3.3$, and $4.4\metre$. At the end of the acceleration, the driver has lost a large fraction of its energy. 
b)~Longitudinal phase space $(p_{||},\xi)$ of the witness at the same positions. 
Initially, the witness develops a longitudinal chirp.  As the witness dephases with the driver, however, the field acting on the witness changes, leading to a reduction in the acquired chirp after $\sim2.8\metre$.
c)~Energy spectra of the witness bunch after an acceleration length of 
$L=1.1$, $2.2$, $3.3$ and $4.4\metre$.  The choice of initial witness parameters results in a near-monoenergetic bunch at the end of the acceleration length.
d)~Evolution of the witness bunch emittance over the entire simulation length.  The large initial growth in emittance is likely numerical in nature.
\label{fig:Evolution}}
\end{figure}

The presence of the co-axial plasma filament, however, stabilizes
the system so that BBU is avoided completely. We follow the acceleration
over a total distance of $L_{\mathrm{acc}}=2\times10^{5}k_{p}^{-1}\approx4.4\metre$.
The phase-space evolution of the driver is shown in Fig.~\ref{fig:Evolution}a.
We observe that at the end of the acceleration length, the driving
bunches have lost $\sim88$\% of their total energy.

The phase-space evolution of the witness bunch is shown in Fig.~\ref{fig:Evolution}b.
The witness initially develops a negative energy chirp, in agreement with the field at $L=0$ shown in Fig.\ref{fig:Accelerating-wakefield}.  However, as the witness is accelerated, it dephases with the drive beam, and so experiences a non-constant accelerating field over the acceleration length.  As we carefully tuned the initial parameters, the gradient of the accelerating field acting on the witness is reversed after an acceleration length of $\sim1\metre$, reducing the chirp.

Figure~\ref{fig:Evolution}c shows the energy spectra of the witness
bunch. The energy spread initially increases due to the chirp, and subsequently decreases, reaching a near-monoenergetic peak at $W\approx21\GeV$ for an acceleration distance of $L=4.4\metre$.  For the witness charge of 410~pC, this corresponds to a total energy of  $8.8\Joules$ -- an energy gain of $8.0\Joules$.  Given the initial driver energy of $20\Joules$ for the 10~nC beam, this represents a 40\% efficiency.  Comparing only the energy lost by the driver gives a transfer efficiency of 46\%.  The measured energy spread $\Delta W=0.2$\% is limited by the simulation resolution.

The normalized emittance of the witness grows rapidly at the start of the simulation from $\epsilon^\ast=1\micron$ to $1.5\micron$, and shows only slight growth over the acceleration length, reaching $1.8\micron$ after $4.4\metre$, as seen in Fig~\ref{fig:Evolution}d.  The rapid growth of the initial emittance is likely numerical in nature.  The resolution-limited emittance at the start of the simulation is estimated at $1.7\micron$ \footnote{Numerical Cherenkov radiation can lead to nonphysical emittance growth
in standard Yee-like PIC codes \protect{\cite{Lehe}}. However, the quasistatic
nature of the qv3d code makes it naturally free from this numerical
artefact. The minimum emittance that can be accurately modelled is
therefore limited only by the simulation resolution. We consider a particle beam focused down to a radius $x_0$, with particles oscillating at the betatron frequency uniformly distributed over all phases. The normalized transverse emittance is then
{$\gamma\sqrt{<x^{2}><{x^{\prime}}^{2}>}\allowbreak=\gamma\sqrt{(x_{0}^{2}/3)(\omega_{\beta}^{2}x_{0}^{2}/c^2)}\allowbreak=(\omega_{\beta}\gamma/\sqrt{3}c)x_{0}^{2}$.}
Choosing a beam diameter of one cell, $x_0=0.5\Delta_{y}$, and noting from simulations that the ion column density at the position of the witness is $\sim6n$ gives the resolution limit for the normalized transverse emittance as $\epsilon^\ast=2.7\times10^{-8}\sqrt{\gamma}$.
}.

The leading edge of the driver expands freely due to its initial emittance, but this leads only to a small perturbation of the wakefield.  For longer propagation distances, this could be compensated with an external focusing field.  Simulations show that the slow pinch of the ion column is vital for beam stabilization - the same configuration with a heavier ion species again results in BBU.  Helium also has the desirable property that no secondary ionization occurs for these parameters.  Dark current injection is avoided, as electrons streaming back from the bulk plasma behind the driver arrive only after the first potential bucket.

The use of plasma as the acceleration medium makes this acceleration scheme extremely flexible.
All lengths scale directly with the plasma wavelength, and so, if desired, a wider channel in a lower-density
plasma can be used to give the same acceleration over a longer propagation distance, which may be easier to achieve experimentally.
The plasma density used here here represents the upper limit for this configuration.  At higher densities, the self-field of the driver becomes sufficient to ionize the bulk gas in the channel.  However, it is also possible to make use of a lower-current driver over a longer propagation distance, which reduces the required beam charge.  In this case, the optimal shape of the driver will be slightly altered due to the nonlinearity of the plasma response.

Due to the use of an ion-column to focus the beam, this mechanism
is only appropriate for an electron driver. However, we note that
this configuration could be used to accelerate a positron witness
bunch with a donut profile. Electrons from the bulk plasma stream back to compensate the ion filament behind the driver, leading to a inversion of the transverse field a short distance $\ll 1/k_p$ from the axis, as can be seen from Fig.~\ref{fig:Channel_configuration}c.  This results in a stable point at which positrons may be accelerated.  Comparing with Fig.~\ref{fig:Channel_configuration}b shows this point coincides with a large positive wakefield. However, the optimization is somewhat more complex than for an electron witness, and so will be discussed in detail in a separate work.

To conclude, we have shown that the use of a co-axial plasma filament within a hollow plasma channel prevents the development
of the beam-breakup instability. 
In the short term, this configuration may serve as a pre-acceleration scheme for a conventional accelerator, increasing the dipole field strength at injection.
Ultimately, though, the ability to stably and efficiently accelerate a witness bunch in a single stage
finally offers a path towards a new generation of novel high-energy
plasma-based accelerators.
The combination of high transformer ratio and monoenergetic
acceleration potentially makes this technology a serious contender
for applications-driven research.

This work has been supported by the Deutsche Forschungsgemeinschaft
and by BMBF.

%

\end{document}